# Dynamic index selection in data warehouses

Stéphane Azefack[1], Kamel Aouiche[2] and Jérôme Darmont[1]

[1]*Université de Lyon (ERIC Lyon 2)*
*5 avenue Pierre Mendès-France*
*69676 Bron Cedex*
*France*
jerome.darmont@univ-lyon2.fr

[2]*Université du Québec à Montréal (LICEF)*
*100 rue Sherbrooke Ouest*
*Montréal (Québec) H2X 3P2*
*Canada*
kamel.aouiche@gmail.com

## Abstract

*Analytical queries defined on data warehouses are complex and use several join operations that are very costly, especially when run on very large data volumes. To improve response times, data warehouse administrators casually use indexing techniques. This task is nevertheless complex and fastidious. In this paper, we present an automatic, dynamic index selection method for data warehouses that is based on incremental frequent itemset mining from a given query workload. The main advantage of this approach is that it helps update the set of selected indexes when workload evolves instead of recreating it from scratch. Preliminary experimental results illustrate the efficiency of this approach, both in terms of performance enhancement and overhead.*

## 1. Introduction

A data warehouse is generally modeled by a star-like schema that contains a central, typically very large fact table, and several dimension tables that describe the facts [1]. An analytical query over such a model necessitates very costly join operations between the fact table and dimension tables.

Selecting suitable physical structures that improve system performance is the role of data warehouse administrators. However, given the wide development of data warehouses, as well as their structural and operational complexity, minimizing the administration function is a crucial issue.

In this context, we have proposed an automatic bitmap join index selection method based on frequent itemset mining from a given query workload [2]. Attributes that frequently appear together in queries indeed constitute good candidate indexes; and bitmap join indexes are particularly appropriate to data warehouses [3]. However, this approach is static: if the input workload significantly evolves with time, we must rerun the whole process to preserve performance.

In this paper, we improve our approach by two aspects. First, we replace the frequent itemset mining technique we used (namely, Close [4]) with an incremental frequent itemset mining technique, so that the selected index configuration can be updated instead of being recreated from scratch. Second, instead of mining closed frequent itemsets, we mine maximal frequent itemsets that are less numerous and help build better indexes. Finally, to the best of our knowledge, this is the first attempt at dynamically selecting indexes in data warehouses.

The remainder of this paper is organized as follows. We present the state of the art regarding both index selection in data warehouses and incremental frequent itemset mining in Section 2. Then, we detail our approach in Section 3 and discuss related experimental results in Section 4. We finally conclude this paper and provide research perspectives in Section 5.

## 2. Related work

### 2.1. Index selection in data warehouses

The index selection problem has been studied for many years in databases, but adaptations to data warehouses are few. In this particular context, research studies may be clustered into two families: algorithms that optimize maintenance cost and algorithms that optimize query response time. In both cases, optimization is realized under storage space constraint. In this paper, we are particularly interested in the second family of approaches, which may be classified with respect to the way a set of candidate indexes and the final configuration of indexes are built.

A set of candidate indexes may be built manually by the administrator, according to his expertise of the workload [5, 6]. This is both subjective and quite hard to achieve when the number of workload queries is very large. In opposition, candidate indexes may also be extracted automatically by syntactically analyzing the workload [7, 8, 9].

There are also several methods for building the final index configuration from candidate indexes. Typically, greedy algorithms increasingly select indexes minimizing workload cost until it does not decrease anymore [5, 6, 7]. Classical optimization algorithms have also been used to solve this problem, such as knapsack resolution [8] and genetic algorithms [10].

## 2.2. Incremental frequent itemset mining

Many algorithms have been proposed in the literature for incrementally mining frequent itemsets. They reuse the frequent itemsets discovered before transaction database update to compute new frequent itemsets. Updating the set of frequent itemsets is very costly, though.

To reduce the problem's dimensionality, closed or maximal frequent itemsets may be mined instead of all frequent itemsets. A frequent itemset $I$ is closed if it contains all the items that occur in every transaction in which $I$ is present. A maximal frequent itemset is a frequent itemset that has no frequent superset.

Most closed frequent itemset mining algorithms exploit concept lattices. The main incremental approach [11] manages lattice updates (unchanged, updated and inserted nodes into the lattice). However, its complexity is quadratic with respect to the number of elements in the concept lattice [12] and the number of closed frequent itemsets may become very large with respect to database size [13].

To the best of our knowledge, the only incremental maximal itemset frequent mining approach, GenMax, exploits a backtracking algorithm to prune the search space as soon as possible with respect to previous iterations [14]. This algorithm also improves support computation and optimizes short term mining.

## 3. Dynamic index selection strategy

In this section, we present the extension of our automatic join index selection method based on frequent itemset mining [2]. Its principle is to exploit transaction logs (i.e., the set of all queries processed by the system) to recommend an index configuration improving data access time.

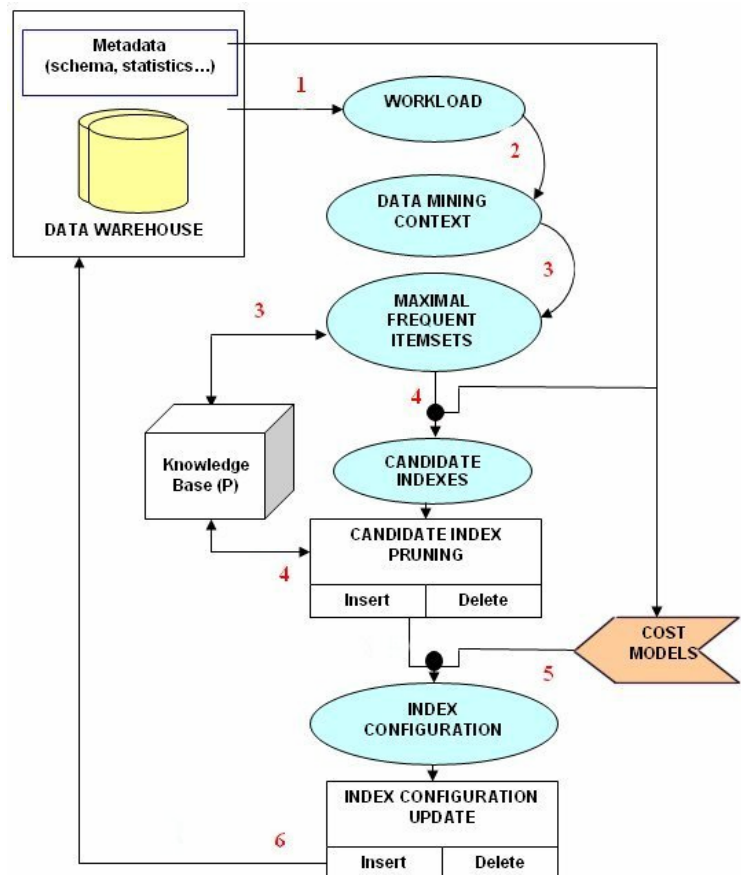

**Figure 1. Dynamic index selection strategy**

This new approach is subdivided into six steps (Figure 1): (1) a workload $Q$ we suppose representative of system usage is extracted from system logs; (2) so-called indexable attributes are extracted from $Q$ and structured in a suitable data mining context $QA$; (3) incremental frequent itemset mining is applied on $QA$, exploiting a knowledge base $P$ that stores information regarding previous executions of this step; (4) emerged (new) frequent itemsets are analyzed to generate new candidate indexes; declined (now infrequent) itemsets correspond to indexes to be dropped; retained (still frequent) itemsets correspond to candidate indexes to retain; the whole set of candidate indexes is labeled $I_C$; (5) since disk space is constrained, $I_C$ is pruned using cost models; (6) the resulting index configuration $I$ is finally effectively updated. The whole process then reiterates after a period of time set by the administrator. We detail its steps in the following sections.

### 3.1. Workload extraction

System workload is typically accessible from the host database management system's transaction log. A given workload $Q$ is supposed representative if it has been measured during a time period the warehouse administrator judges sufficient to anticipate upcoming transactions.

Since we are more particularly interested in analytical query performance and not warehouse maintenance, we only consider interrogation query workloads in this paper. These queries are typically composed of join operations between the fact table and dimensions, restriction predicates, and aggregation and grouping operations. More formally, an analytic query $q$ may be expressed as follows in relational algebra: $q = \pi_{G, M} \sigma_R (F \bowtie D_1 \bowtie \ldots \bowtie D_n)$; where $G$ is the set of attributes from dimensions $D_1, \ldots, D_n$ that are present in $q$'s grouping clause, $M$ is a set of aggregate measures from fact table $F$ and $R$ a conjunction of predicates over dimension attributes.

### 3.2. Workload analysis

Attributes $a_j$ that may support indexes belong to the sets $G$ and $R$ defined in Section 3.1 [7, 8, 9]. We reference them in a "query-attribute" binary matrix $QA$ whose rows represent workload queries $q_i \in Q$ and whose columns are indexable attributes $a_j$. The general term $QA_{ij}$ of this matrix is equal to one if attribute $a_j$ is present in query $q_i$, and to zero otherwise. This data structure or extraction context directly corresponds to attribute-value tables that are exploited by data mining algorithms.

### 3.3. Maximal frequent itemset mining

In the static version of our approach, we have used the Close closed frequent itemset mining algorithm [4] to obtain a set of candidate indexes $I_C$. In this dynamic extension, we replace it by the GenMax incremental, maximal frequent itemset mining algorithm [14].

In our context, workload $Q$ can be very large and evolves with time. We selected GenMax because it can determine, in a short time, all maximal frequent itemsets from large databases, by optimizing support computation and infrequent itemset pruning through a backtracking process. Moreover, queried data from $Q$ are typically correlated, which leads to a dense extraction context $QA$. Since incremental frequent itemset mining may produce a number of closed frequent itemsets exponentially greater than the number of maximal frequent itemsets [13], GenMax helps produce a smaller quantity of candidate indexes, which reduces the dimensionality of index selection and improves scalability.

In summary, GenMax, in a first iteration, exploits an input transaction database $D$ to produce a knowledge base $P$ that stores, e.g., the list of maximal frequent itemsets from $D$, non-maximal, but frequent itemsets, the number of transactions in $D$, etc. $P$ is then exploited and updated in the next iterations. At each of these iterations, the list of new transactions $d^+$ and the list of deleted transactions $d^-$ are used to compute the updated transaction database $\Delta = (D \cup d^+) - d^-$. Frequent itemset computation is then performed on $\Delta$, using $P$, to minimize accesses to $D$.

### 3.4. Candidate indexes generation

The application of GenMax onto matrix $QA$ helps obtain: a set $I^+$ of emerged frequent itemsets, which were infrequent in $P$ but become frequent in $\Delta$; a set $I^-$ of declined frequent itemsets, which were frequent in $P$ but become infrequent in $\Delta$; and a set $I^0$ of retained frequent itemsets, which are frequent in both $P$ and $\Delta$. Then, the set of candidate indexes is $I_C = (I \cup I^+) - I^-$, where $I$ is the current index configuration. Note that $I^0$ is not used to compute $I_C$, but is nonetheless recorded in $P$.

### 3.5. Candidate indexes selection

The number of candidate indexes in $I_C$ is generally proportional to the size of workload $Q$. Thus, it is not feasible to build all the proposed indexes because of system limitations (e.g., a limited number of indexes per table) or storage space constraints. To circumvent these limitations, we exploit cost models that help greedily select the most advantageous indexes. These models estimate storage space occupied by bitmap join indexes, data access cost whether using these indexes or not, and index maintenance cost. Due to space constraints, we cannot elaborate on these cost models in this paper, but the interested reader can refer to [2] for complete details.

### 3.6. Index configuration update

Applying index selection (Section 3.5) on $I_C$ outputs a new index configuration $I'$. To update the current index configuration $I$, we must eventually: create all indexes $i \in I'$ such that $i \notin I$, i.e., all emerged indexes

$i \in (I' - I)$; drop all indexes $i \in I$ such that $i \notin I'$, i.e., all declined indexes $i \in (I - I')$; and reset $I$ to $I'$.

## 4. Experiments

### 4.1. Experimental conditions

To illustrate the advantage of our dynamic index selection approach over our static one, we ran tests on a 1 GB data warehouse implemented within Oracle 9i, on a Pentium Dual Core 1.6 GHz PC with 1 GB RAM and a 120 GB IDE disk under Windows XP Pro. Our test data warehouse is derived from Oracle's, whose classical sales star schema is composed of one fact table and five dimensions. We have defined an initial workload $Q1$ of 30 analytical queries involving aggregation operations and multiple joins between the fact table and dimension tables. Then, we defined four evolutions of $Q1$ ($Q2$ to $Q5$), so that some frequent attribute sets emerge, some decline, and others remain frequent. Due to space constraints, we reproduce here neither the full data warehouse schema nor the detail of each workload, but they are available on demand.

### 4.2. Results

We ran our tests for each workload $Q1$ to $Q5$; without indexing (for reference), with static and dynamic indexing; and for an arbitrary minimum support value (when mining frequent itemsets) of 0.05 that is low enough to produce a fair number of candidate indexes. In each of these tests, we measured index selection time (with both our static and dynamic approaches), index creation (static approach) or update (dynamic approach) time under Oracle, and workload execution time. The results we obtained are plotted in Figures 2, 3 and 4, respectively. All three figures feature workloads ($Q1$ to $Q5$) on the X axis and execution times on the Y axis (in milliseconds, seconds and minutes, respectively).

Figure 2 shows that the overhead of our dynamic approach, in term of index selection alone, is about 5.2 times greater than that of our static approach, on an average. However, this is not due to the algorithms' intrinsic complexity, but to our implementations. The static approach has been implemented and optimized in PHP, while our newer, dynamic approach is implemented in Java (which appears slower than PHP on a standalone workstation) and not optimized yet. Both their execution times remain in the same order of magnitude (hundreds of milliseconds here), though.

On the other hand, Figure 3 shows that index update time is about 9.5 faster with our dynamic approach on an average, and even about 11.5 times faster if $Q1$'s execution (and thus initial index configuration creation) is excluded from computation. Since index update runs in tens of seconds in our examples, the main overhead of our method lies here, and the enhancement brought by dynamicity is obvious.

Figure 4 eventually shows that exploiting maximal frequent itemsets (dynamic approach) instead of closed frequent itemsets (static approach) helps select more pertinent indexes, since response time is slightly better (about 13% on an average) in the dynamic case. This is presumably because fewer candidate indexes are generated and then selected, which simplifies index choice at query optimization time.

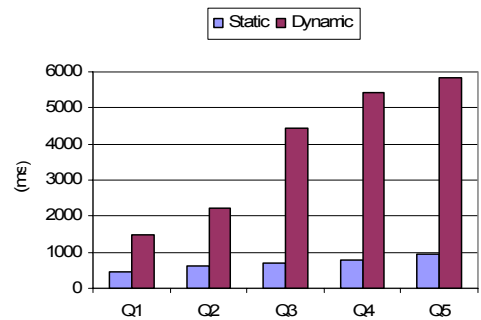

**Figure 2. Index selection time**

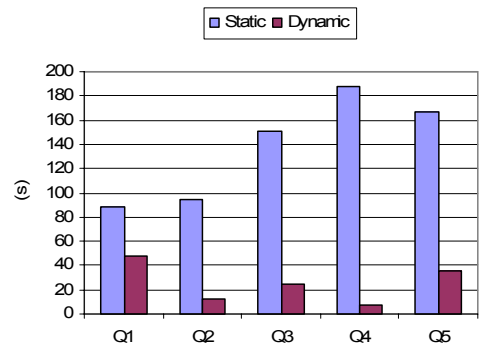

**Figure 3. Index creation/update time**

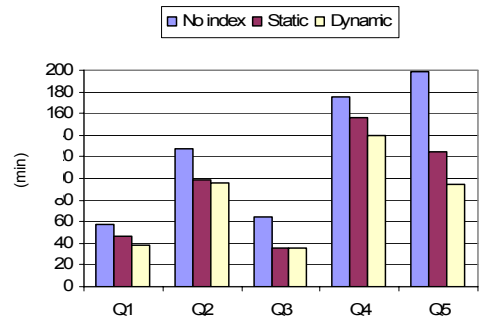

**Figure 4. Workload execution time**

## 5. Conclusion and perspectives

In this paper, we have presented a dynamic bitmap join index selection method for data warehouses that is based on incremental frequent itemset mining from a given query workload. The main advantage of this approach is that it helps update the set of selected indexes when workload evolves instead of recreating it from scratch.

Our first experiments (which we aim to extend, notably by scaling them up, to complete our approach's validation) indeed show that introducing dynamicity helps reduce index maintenance overhead. Furthermore, exploiting maximal instead of closed frequent itemsets also helps improve the index configuration's quality, and hence query response time.

Furthermore, note that our approach is purposely modular and generic. Each step (frequent itemset mining, candidate indexes selection…) exploits interchangeable tools. The data mining technique and cost models we use are indeed not related to any system in particular and could easily be replaced by other, more efficient methods if necessary.

Eventually, a critical issue when using automatic, dynamic optimization strategies is to master system overhead, and in particular determine when the administrator should run the incremental index update process. Pursuing this lead is our main research perspective. Studies related to session detection that are based on entropy computation [15] could be very useful for this sake.